\documentclass[12pt,italian]{article}

\usepackage{graphicx}

\title{\bf Analisi dimensionale: due interessanti applicazioni}
\author{Germano D'Abramo\\
{\small Istituto di Astrofisica Spaziale e Fisica Cosmica,}\\
{\small Area di Ricerca CNR Tor Vergata, Roma, Italy}\\
{\small E--mail: {\tt dabramo@rm.iasf.cnr.it}}}

\date{\small 5 Giugno 2003}

\begin{document}

\maketitle

\section{Introduzione}

Malgrado non sia molto conosciuta persino fra gli addetti ai lavori,
l'{\em analisi dimensionale} costituisce uno degli strumenti pi\`u potenti
e intriganti delle scienze fisiche.

Ma in cosa consiste? Brevemente, dato un sistema fisico complesso, come ad
esempio la formazione di un'onda d'urto nucleare o la creazione di un
cratere d'impatto, e definite tutte le variabili e le costanti fisiche che
sembrano descrivere propriamente e completamente il processo, spesso
risulta possibile scrivere in maniera univoca delle leggi matematiche
(solitamente leggi di potenza) che forniscono la giusta dipendenza
funzionale (dal punto di vista fisico)  di una di queste variabili
rispetto a tutte le altre. E questo pu\`o essere fatto semplicemente
prendendo in considerazione le dimensioni fisiche fondamentali delle
variabili e costanti fisiche coinvolte: lunghezza, massa, tempo, carica
elettrica (in alcuni casi specifici anche la temperatura) e le loro
combinazioni.

Un'applicazione immediata. Supponiamo di non sapere molto di cinematica e
di voler ottenere la relazione matematica che lega l'accelerazione
centripeta, $a$, la frequenza di rotazione, $\omega$\footnote{La frequenza
di rotazione \`e definita come $2\pi /T$, dove $T$ \`e il periodo di
rotazione.}, e la distanza dal centro, $r$, di un corpo che ruota a
velocit\`a uniforme lungo una traiettoria perfettamente circolare.
Euristicamente parlando (e anche in base alla nostra esperienza diretta),
\`e ragionevole immaginare che $a$ dipenda proprio solo da $\omega$ e $r$.

Consideriamo allora le dimensioni fisiche di $a,\omega$ e $r$:

\begin{equation}
\begin{array}{lcl}
[a] & = & L\cdot T^{-2},\\
{[\omega]} & = & T^{-1},\\
{[r]} & =  & L.
\end{array}
\label{eq1}
\end{equation}

Il segno $[x]$ sta per ``le dimensioni fisiche di $x$'' e $L$ sta per 
unit\`a di {\em lunghezza} e $T$ per unit\`a di {\em tempo}.

Ora, l'unico modo di ottenere le dimensioni fisiche di $a$ \`e di 
combinare $\omega$ e $r$ nella maniera che segue:

\begin{equation}  
\omega^2 r.
\label{eq2}
\end{equation}   

Va notato che la quantit\`a 
\begin{equation}
\frac {a}{\omega^2r},
\label{eq3}
\end{equation}
\`e adimensionale e che quindi risulta invariante per cambiamento di 
unit\`a di misura, ad esempio passando da metri a chilometri, da secondi a 
ore.

Quindi, abbiamo che la~(\ref{eq3}) \`e una costante
\begin{equation}  
\frac {a}{\omega^2r}=k,
\label{eq4}
\end{equation}   
cio\`e

\begin{equation}
a= k {\omega^2r},
\label{eq5}
\end{equation}
che \`e proprio la dipendenza funzionale che si otterrebbe sviluppando
matematicamente la teoria cinematica del moto circolare uniforme.

Ovviamente, l'analisi dimensionale pu\`o fornire la dipendenza funzionale
tra le variabili ma non \`e capace di definire le costanti moltiplicative
adimensionali, la costante $k$ nel nostro caso. Per queste sono
indispensabili o una vera teoria fisica o un confronto diretto fra le
relazioni ottenute e i dati sperimentali. Va detto inoltre, come vedremo
in seguito, che tra i limiti di questo metodo di studio c'\`e che non
sempre \`e possibile definire univocamente la relazione funzionale tra le
variabili coinvolte e anche in questo caso \`e necessario, in mancanza di
una vera teoria, un confronto diretto con i dati sperimentali. Per inciso,
nel semplice caso analizzato sopra vale esattamente $a=\omega^2 r$ (cio\`e
la costante adimensionale $k$ \`e proprio pari a 1).

Pi\`u in generale, supponiamo di avere una relazione funzionale tra una
quantit\`a $a$, che deve essere determinata in un esperimento, e un
insieme di parametri $(a_1,...,a_{n-1})$, che sono sotto il controllo
sperimentale. Sia $a$ che l'insieme $(a_1,...,a_{n-1})$ possiedono
dimensioni fisiche. La forma generale della relazione pu\`o essere la
seguente

\begin{equation}
a=f(a_1,...,a_{n-1}).
\label{eq5b}
\end{equation}

Ora, se il minimo numero di parametri nell'insieme $(a,a_1,...,a_{n-1})$
con i quali posso ottenere tutte le dimensioni fisiche presenti nel mio
problema \`e $k<n$, il Teorema~$\Pi$ di Buckingham~\cite{Bu}, che
rappresenta il risultato centrale dell'analisi dimensionale, afferma che
il sistema fisico allo studio dipende da un insieme di $n-k$ parametri
adimensionali $(\Pi_1,...,\Pi_{n-k})$. Questi parametri sono combinazioni
adimensionali indipendenti delle quantit\`a $(a,a_1,...,a_{n-1})$; essi
sono indipendenti nel senso che, per ogni $i$, non \`e possibile ottenere
$\Pi_i$ da una combinazione di altri $\Pi$s.

In questo modo, l'eq.~(\ref{eq5b}) pu\`o essere scritta come

\begin{equation}
\Pi_1=\Phi(\Pi_2,...,\Pi_{n-k}),
\label{eq5c}
\end{equation}
dove $\Phi$ \`e una funzione adimensionale generale. Ancora, per ottenere 
la forma esatta di $\Phi$ abbiamo bisogno di una vera teoria del fenomeno 
fisico allo studio, o un confronto diretto con i dati sperimantali.

In questo articolo applichiamo le procedure dell'analisi dimensionale a
due casi molto diversi fra loro (e con finalit\`a diverse), ma entrambi
piuttosto interessanti.

Nel primo caso (sezione {\bf 2}) si usa l'analisi dimensionale non per
derivare una relazione funzionale tra quantit\`a fisiche, ma per capire
come cambia la scala temporale dei fenomeni fisici al cambiare del valore
numerico dei parametri importanti, quali la gravit\`a superficiale di un
pianeta.

Nel secondo caso (sezione {\bf 3}), invece, utilizziamo l'analisi
dimensionale nella maniera pi\`u canonica per ottenere una relazione
matematica che leghi l'ampiezza di un onda marina alla sua velocit\`a.
Applicazione immediata: la fisica degli {\em tsunami}\footnote{Onde
anomale particolarmente distruttive. Letteralmente {\em onde di porto},
dal giapponense {\em tsu} (porto) e {\em nami} (onda).}.

\section{Moviola sulla Luna}

Avete mai visto, proiettati a velocit\`a doppia, i filmati degli
astronauti che si muovono sulla superficie lunare? Se provate a farlo vi
accorgerete sicuramente di una caratteristica stupefacente: sembrano
girati sulla Terra. I movimenti degli astronauti, i sobbalzi del rover
motorizzato etc.~sembrano proprio quelli naturali che si avrebbero se le
stesse operazioni venissero compiute sulla Terra.

\`E questo un ulteriore elemento a favore della famigerata teoria secondo
la quale la NASA non avrebbe mai portato l'uomo sulla Luna?

Molto probabilmente no. In ogni caso qui di seguito mostriamo come questo
curioso comportamento sia legato semplicemente alla gravit\`a superficiale
dei pianeti.

In generale, ogni fenomeno dinamico la cui evoluzione nel tempo dipenda
anche dalla gravit\`a di un pianeta (ad esempio, un essere umano che
cammina e salta, un palazzo che crolla sotto il proprio peso, una slavina
o una colata di magma che scivolano a valle) ha un {\em tempo
caratteristico} che dipende proprio dalla gravit\`a superficiale. Il
tempo caratteristico non \`e una costante fisica o un numero ben preciso,
ma \`e un concetto generale. Per dare un idea, il tempo
caratteristico della vita umana \`e circa 80 anni, mentre quello della
dinamica orbitale dei pianeti interni del Sistema Solare \`e dell'ordine
dell'anno (periodo di rotazione della Terra intorno al Sole).

Ad esempio, una delle pi\`u semplici equazioni della cinematica ci dice
che una biglia che cada da ferma in un campo gravitazionale costate $g$
percorre nel tempo $t$ uno spazio pari a

\begin{equation}
x=\frac{1}{2}gt^2.
\label{eq6}   
\end{equation}

Invertendo l'equazione~(\ref{eq6}), si ottiene il tempo in funzione dello
spazio percorso

\begin{equation}
t=\sqrt{\frac{2x}{g}}.
\label{eq7}   
\end{equation}

L'equazione~(\ref{eq7}) ci mostra proprio la dipendenza del tempo
caratteristico dalla costante gravitazionale $g$. Infatti, pi\`u grande
\`e $g$ e minore \`e il tempo impiegato dalla biglia per percorrere in
caduta libera lo spazio $x$.

In generale, quindi, i tempi caratteristici con cui si verificano i
fenomeni dinamici sulle superfici planetarie con gravit\`a superficiali
$g_1$ e $g_2$ scalano come

\begin{equation}
\frac{t_1}{t_2}=\sqrt{\frac{g_2 2x}{g_1 2x}},\qquad\textrm{cio\`e}\qquad
T_1=T_2\sqrt{\frac{g_2}{g_1}}.
\label{eq8}   
\end{equation}

Nel caso specifico dei filmati lunari il tempo caratteristico lunare
$T_{\textrm{\tiny\bf )}}$ \`e pari a

\begin{equation}
T_{\textrm{\tiny\bf )}}=T_\oplus\sqrt{\frac{g_\oplus}{g_{\textrm{\tiny\bf 
)}}}}=T_\oplus\sqrt{\frac{9.78}{1.62}}\simeq 2.45 T_{\oplus}, 
\label{eq9}
\end{equation}
dove $T_\oplus$, $g_\oplus$ e $g_\textrm{\tiny\bf )}$ sono rispettivamente
il tempo caratteristico terrestre, la costante di gravit\`a superficiale
terrestre e quella lunare (ricordiamo che la gravit\`a superficiale lunare
\`e circa 1/6 di quella terrestre).

Cio\`e, $T_{\textrm{\tiny\bf )}}$ \`e poco pi\`u di due volte maggiore del
tempo caratteristico dei fenomeni dinamici sulla superficie terrestre. E
quindi, se aumentiamo di due volte la velocit\`a di proiezione dei filmati
lunari otteniamo proprio l'impressione che i fenomeni dinamici si svolgano
con la stessa tempistica dei fenomeni terrestri! Ricordiamo per\`o che la
relazione~(\ref{eq9}) vale solo per i fenomeni che sono soggetti alla
forza di gravit\`a: se fosse possibile intravvederle, noteremmo che le
palpebre degli astronauti battono ad una velocit\`a innaturalmente doppia!

\section{Cavalcare lo tsunami}

In questa sezione, sempre grazie all'ausilio dell'analisi dimensionale,
deriveremo la relazione che lega l'ampiezza di un onda marina alla sua
velocit\`a.

Attraverso questa relazione si pu\`o comprendere per linee generali il
motivo fisico della formazione degli tsunami, ovvero delle tanto temute
{\em onde anomale} che possono raggiungere altezze ragguardevoli in
prossimit\`a della riva ed essere altamente distruttive.

Cerchiamo in primo luogo di ottenere l'energia totale che pu\`o essere
tras\-por\-ta\-ta da un'onda marina. Come sar\`a chiaro nello sviluppo
successivo, l'uso dell'analisi dimensionale ci esime dalla necessit\`a di
rappresentare (e semplificare) l'onda marina come {\em sinusoidale piana},
cio\`e periodica, il cui profilo segue perfettamente la funzione
trigonometrica seno (o coseno) e il cui fronte \`e una lina retta {\em
infinita}.

Visto che abbiamo a che fare con un fenomeno fisico che coinvolge l'acqua
del mare, cominciamo ad elencare tutte le quantit\`a fisiche che
euristicamente ci aspettiamo possano entrare in gioco nella descrizione
del fenomeno allo studio, insieme con le loro unit\`a di misura.

Immaginiamo di essere interessati per prima cosa all'energia cinetica
tras\-por\-ta\-ta dall'onda per unit\`a di superficie attraversata (che
chiamiamo $E_c/S$). Se parliamo di energia cinetica non possiamo allora
non considerare la velocit\`a dell'onda, $v$, che sicuramente influisce,
cos\`{\i} come la sua ampiezza, $A$, e la densit\`a del mezzo in cui
l'onda si propaga, $\rho$, che fanno le veci della massa nella classica
relazione dell'energia cinetica $E=\frac12mv^2$. \`E ovvio che in questo
caso non possiamo utilizzare $m$ perch\'e non stiamo parlando di un
oggetto dai contorni fisici ben definiti.

\begin{equation}
\frac{E_c}{S},\,\,\, \rho,\,\,\,  v,\,\,\, A,\,\,\, \lambda.
\label{eq10}
\end{equation}


\begin{figure}[t]
\centerline{\includegraphics[width=8cm,angle=0]{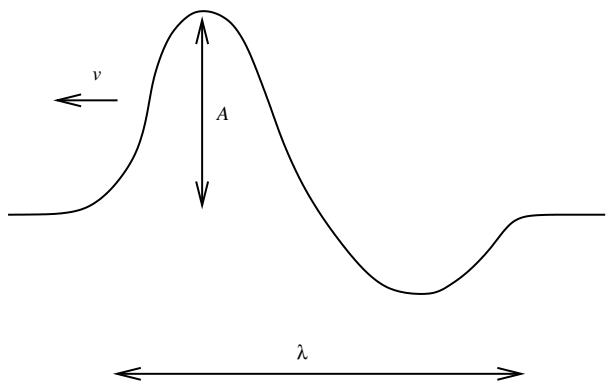}}
\caption{Rappresentazione schematica di un'onda, con le grandezze fisiche 
che la definsicono (per ovvi motivi non \`e rappresentata la densit\`a 
$\rho$).}
\label{fig1}
\end{figure}

La lista~(\ref{eq10}) raccoglie quindi tutte le quantit\`a fisiche
considerate importanti. In realt\`a compare anche $\lambda$, che \`e la
lunghezza d'onda dell'onda marina. Essa \`e una caratteristica geometrica
dell'onda che non abbiamo ragione di escludere. 

Per la derivazione di $E_c/S$ non abbiamo preso in considerazione la
costante di accelerazione gravitazionale $g$. Ci aspettiamo infatti che
per $E_c/S$ la gravit\`a non sia rilevante e che l'energia cinetica sia
determinata una volta fissate $A$, $\rho$, $v$ e $\lambda$. Prenderemo in
considerazione $g$ pi\`u avanti quando deriveremo l'energia potenziale
(gravitazionale) dell'onda.

Veniamo ora alle unit\`a di misura (dimensioni) dei parametri 
in~(\ref{eq10}): 

\begin{equation}
\begin{array}{lcl}
[E_c/S] & = & M\cdot T^{-2},\\
{[\rho]} & = & M \cdot L^{-3},\\
{[v]} & = & L\cdot T^{-1},\\
{[A]} & = & L,\\
{[\lambda]} & = & L.
\end{array}   
\label{eq11}
\end{equation}
dove $M$ sta per unit\`a di {\em massa}, $T$ unit\`a di {\em tempo} e $L$ 
unit\`a di {\em lunghezza}.

Dall'elenco~(\ref{eq11}) si pu\`o notare come con i parametri $A$, $\rho$
e $v$ \`e possibile costruire, attraverso le giuste combinazioni, le
dimensioni di tutti gli altri. Essi sono cio\`e i parametri dimensionali
indipendenti del problema secondo la teoria dell'analisi dimensionale di
Buckingham vista nell'introduzione. Sempre secondo la teoria, si ha che ci
sono allora $5-3=2$ paramentri adimensionali

\begin{equation}
\Pi_1 = \frac{E_c/S}{\rho v^2 A},\qquad \Pi_2 = \frac{\lambda}{A},
\label{eq11b}  
\end{equation}
e quindi, per il Teorema~$\Pi$, la relazione funzionale generale che lega
$E_c/S$ a $A$, $\rho$, $v$ e $\lambda$ \`e

\begin{equation}
\frac{E_c}{S} = \rho v^2 A \Phi\Biggl(\frac{\lambda}{A}\Biggr),
\label{eq12} 
\end{equation}
dove $\Phi$ \`e una funzione adimensionale indeterminata, per ottenere 
la quale \`e necessaria una vera teoria o un confronto diretto con i dati 
sperimentali.

Ma qual'\`e il comportamento della funzione $A\Phi (\lambda /A)$ al
variare di $A$ e $\lambda$? Euristicamente ci aspettiamo che se $A$ e/o
$\lambda$ aumentano, deve aumentare anche $\frac{E_c}{S}$. Infatti, se $A$
e/o $\lambda$ aumentano a parit\`a di $v$ e $\rho$, le dimensioni fisiche
dell'onda aumentano e con loro la sua {\em massa} e dunque la sua energia
cinetica. Quindi, la funzione $\Phi (\lambda /A)$ deve essere
proporzionale al rapporto $\lambda /A$ per rendere conto dell'aumento di
$\Phi$ all'aumentare di $\lambda$ (con $A$ fissato).  Inoltre, $A\Phi
(\lambda /A)$ deve crescere all'aumentare di $A$, con $\lambda$ costante.
Ne consegue che all'aumentare di $A$, $\Phi (\lambda /A)$ deve diminuire
(e diminuisce poich\'e il rapporto $\lambda /A$ diminuisce) pi\`u
lentamente di $A$ per ogni valore di $\lambda$. Un semplice esempio di
funzione $\Phi$ che soddisfa queste caratteristiche potrebbe essere il
seguente

\begin{equation}
\frac{E_c}{S} = k_c \rho v^2 A^{\alpha}\lambda^{\beta},
\label{eq12b}   
\end{equation}
dove $k_c$ \`e una costante adimensionale sconosciuta, $\alpha$ e $\beta$
sono due numeri qualunque ma maggiori di zero e tali che $\alpha +
\beta=1$.

Come anticipato poco f\`a, l'onda marina trasporta anche energia
potenziale gravitazionale, per il semplice fatto che una certa massa
d'acqua, il picco dell'onda, \`e sollevata rispetto al pelo libero della
superficie del mare. Cerchiamo allora di derivare euristicamente l'energia
potenziale dell'onda che attraversa l'unit\`a di superficie, $E_p/S$. Come
prima, le quantit\`a fisiche da cui $E_p/S$ pu\`o dipendere sono la
densit\`a $\rho$ dell'acqua, che mutua la {\em massa dell'onda},
l'ampiezza $A$ e la lunghezza d'onda $\lambda$ che esprimono la grandezza
(altezza e lunghezza) dell'onda con cui abbiamo a che fare e infine la
costante di acceleraione gravitazionale $g$, che tiene conto della forza
di gravit\`a rispetto alla quale l'onda possiede un'energia potenziale.

Passando alle unit\`a di misura (dimensioni), si ha:

\begin{equation}
\begin{array}{lcl}
[E_p/S] & = & M\cdot T^{-2},\\
{[\rho]} & = & M \cdot L^{-3},\\
{[A]} & = & L,\\
{[\lambda]} & = & L,\\
{[g]} & = & L\cdot T^{-2}.
\end{array}
\label{eq13}
\end{equation}

Anche qui, scegliendo come parametri indipendenti $A$, $\rho$ e $g$, si ha
che ci sono $5-3=2$ paramentri adimensionali

\begin{equation}
\Pi_1 = \frac{E_p/S}{\rho g A^2},\qquad \Pi_2 = \frac{\lambda}{A},
\label{eq13b}
\end{equation}
e quindi, sempre secondo il Teorema~$\Pi$, la relazione funzionale
generale che lega $E_p/S$ a $A$, $\rho$, $g$ e $\lambda$ \`e

\begin{equation}
\frac{E_c}{S} = \rho g A^2 \Psi\Biggl(\frac{\lambda}{A}\Biggr),
\label{eq14}
\end{equation}
dove $\Psi$ \`e una funzione adimensionale indeterminata, per ottenere la
quale, di nuovo, \`e necessaria una vera teoria o un confronto diretto con
i dati sperimentali. Anche in questo caso, il comportamento della funzione
$A^2 \Psi (\lambda / A)$ \`e molto simile a quello della funzione $A\Phi
(\lambda /A)$. Infatti, al crescere di $A$ e/o $\lambda$ l'onda cresce in
dimensioni e pi\`u l'onda \`e grande, maggiore \`e la sua energia
potenziale, cio\`e $\frac{E_p}{S}$.

Quindi, come prima, la funzione $\Psi (\lambda /A)$ deve essere
proporzionale al rapporto $\lambda /A$ per rendere conto dell'aumento di
$\Psi$ all'aumentare di $\lambda$ (con $A$ fissato). Ma $A^2\Psi (\lambda
/A)$ deve crescere all'aumentare di $A$, con $\lambda$ costante: ne
consegue che all'aumentare di $A$, $\Psi (\lambda /A)$ deve diminuire
pi\`u lentamente di $A^2$ per ogni valore di $\lambda$. Come prima, un
semplice esempio di funzione $\Psi$ che soddisfa queste caratteristiche
potrebbe essere il seguente

\begin{equation}
\frac{E_p}{S}= k_p \rho g A^{\gamma} \lambda^{\delta},
\label{eq14b}
\end{equation}
dove, ancora, $k_p$ \`e una costante adimensionale sconosciuta e $\gamma$
e $\delta$ due numeri positivi tali che $\gamma + \delta=2$.

Ma ora torniamo agli tsunami. Se trascuriamo le dissipazioni legate
all'attrito interno dell'acqua e con la sabbia della riva poco prima
dell'impatto dell'onda, possiamo considerare l'energia totale dell'onda
che attraversa l'unit\`a di superficie, $E_t/S$, come una quantit\`a quasi
costante durante la fase di formazione dello tsunami, cio\`e

\begin{equation}
\frac{E_t}{S}=\frac{E_c}{S}+\frac{E_p}{S} \simeq K_t,
\label{eq15}
\end{equation}
per cui la relazione funzionale che ci fornisce $\frac{E_t}{S}$ \`e

\begin{equation}
\rho v^2 A \Phi\Biggl(\frac{\lambda}{A}\Biggr)  + \rho g A^2 
\Psi\Biggl(\frac{\lambda}{A}\Biggr) \simeq K_t.
\label{eq16}
\end{equation}

A questo punto, nei pressi della spiaggia $v$ deve necessariamente
diminuire (si avvicina ad una zona dove non c'\`e pi\`u acqua!), cos\`{\i}
come deve ridursi la lunghezza d'onda $\lambda$: in un certo senso l'onda
marina rallenta e si comprime in prossimit\`a della riva.
Dall'equazione~(\ref{eq16}) risulta allora chiaro che se $v$ e $\lambda$
diminuiscono, affinch\'e $E_t/S$ rimanga quasi costante \`e necessario che
$A$ aumenti! Non c'\`e altra possibilit\`a di soddisfare tale vincolo,
visti i comportamenti di $A\Phi (\lambda /A)$ e $A^2\Psi (\lambda /A)$
analizzati pi\`u sopra; per ogni valore istantaneo di $\lambda$, l'aumento
di $A$ implica l'aumento di $A\Phi (\lambda /A)$ e $A^2\Psi (\lambda /A)$.

Per completezza va detto che la relazione~(\ref{eq16}) \`e stata ottenuta
non tenendo conto dell'influenza che potrebbe avere la profondit\`a del
fondale (nelle equazioni~(\ref{eq12}) e (\ref{eq14}) la profondit\`a $h$
non compare): questo ha reso la derivazione della~(\ref{eq16})  alquanto
diretta, ma ha anche confinato il {\em range} di applicabilit\`a in quella
che pu\`o essere chiamata {\em zona asintotica}, quella zona dove il
fondale non influisce poich\'e ha una profondit\`a il cui valore numerico
\`e molto pi\`u grande di quello dell'ampiezza dell'onda ($h\gg A$).

Ma allora dobbiamo credere o no alla dipendenza funzionale di $A$ da $v$
in prossimit\`a della riva mostrata dalla~(\ref{eq16})? La risposta
rigorosa \`e no, tuttavia se $\lambda$ \`e originariamente (ovvero quando
$h\gg A$) pi\`u grande della lunghezza del tratto di mare
in cui il fondale passa da $h\gg A$ a $h\simeq A$, allora possiamo dire
che la relazione~(\ref{eq16}) \`e accettabile.

In ogni caso, la~(\ref{eq16}) ci mostra indicativamente qual'\`e la 
tendenza di $A$ in funzione di $v$ quando $v$ diminuisce in prossimit\`a 
della riva.

A questo punto pu\`o essere utile fornire qualche informazione relativa a
tsunami verificatisi realmente in passato~\cite{phytsu}. Molti degli
tsunami di cui si ha documentazione storica (anche fotografica, vedi
fig.~\ref{fig3})  hanno avuto origine nell'Oceano Pacifico. Nell'Oceano
Pacifico, dove la profondit\`a media delle acque \`e circa di 4000 metri,
uno tsunami viaggia ad una velocit\`a compresa tra i 700 e i 1000
chilometri orari (la velocit\`a di crociera di un jet di linea). Inoltre,
la sua ampiezza \`e spesso minore di un metro, mentre la sua lunghezza
d'onda \`e mediamente di 100 chilometri (il passaggio di un'onda di
tsunami in alto mare non pu\`o essere praticamente percepita). In
prossimit\`a della riva, tuttavia, l'ampiezza dell'onda sale mediamente
fino ad un massimo di 30 metri!



\begin{figure}[t]
\centerline{\includegraphics[width=8cm,angle=0]{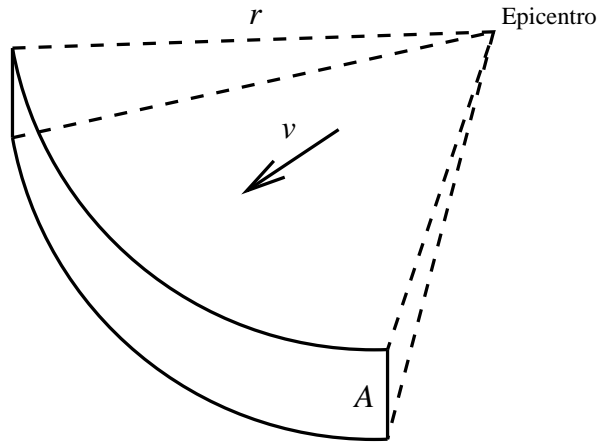}}
\caption{Esemplificazione grafica del fattore di attenuazione geometrica 
dell'energia dell'onda (eq.~\ref{eq17}). }
\label{fig2}
\end{figure}

Passiamo ora a occuparci brevemente della quantit\`a $K_t$, l'energia
totale dell'onda marina per unit\`a di superficie attraversata. Supponiamo
che un'onda anomala sia generata in alto mare da un evento naturale di
grandi proporzioni (terremoto, impatto asteroidale, improvvise
modificazioni orografiche o eruzioni vulcaniche subacquee, etc.) e che
l'energia trasferita al moto dell'acqua sia $E$ (che sar\`a solo una
frazione dell'energia totale sprigionata dall'evento). Ora, se il tratto
di mare in cui si verifica il passaggio da $h\gg A$ a $h\simeq A$ (e in
cui c'\`e la riva) dista $r$ dall'epicentro del fenomeno, allora possiamo
supporre che $K_t$ abbia il seguente andamento funzionale

\begin{equation}
K_t \propto \frac{E}{rA_{h\gg A}},
\label{eq17}
\end{equation}
dove il fattore $\frac{1}{rA_{h\gg A}}$ tiene conto dell'attenuazione
dovuta alla geometria: l'energia si distribuisce approssimativamente sul
bordo di una circonferenza di raggio $r$ e per un'altezza dell'ordine di
$A_{h\gg A}$ (che \`e l'ampiezza dell'onda anomala prima che incontri la
riva e si trasformi in tsunami), e quindi la superficie su cui si
distribuisce \`e in qualche modo proporzionale a $rA_{h\gg A}$.


\begin{figure}[t]
\centerline{\includegraphics[width=13cm,angle=0]{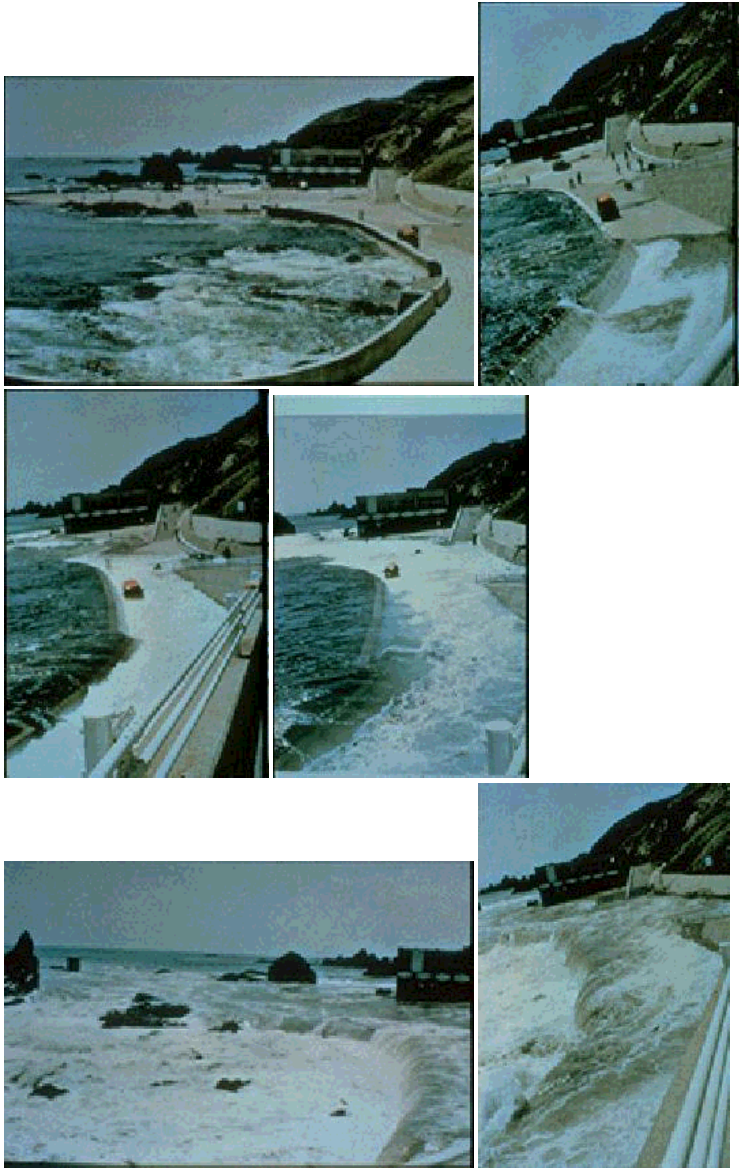}}
\caption{Costa giapponese, 1983.}
\label{fig3}
\end{figure}

\end{document}